# Ultrafast all-optical switching enabled by epsilon-near-zero modes in metal-insulator nanocavities

Joel Kuttruff[2], Denis Garoli[3], Jonas Allerbeck[2], Roman Krahne[3], Antonio De Luca[4], Daniele Brida[1,2], Vincenzo Caligiuri[4,3]*, and Nicolò Maccaferri[1]*

[1]Department of Physics and Materials Science, University of Luxembourg, 1511 Luxembourg, Luxembourg

[2]Department of Physics and Center for Applied Photonics, University of Konstanz, 78457 Konstanz, Germany

[3]Istituto Italiano di Tecnologia, Via Morego 30, 16163 Genova, Italy

[4]Dipartimento di Fisica, Università della Calabria, 87036 Rende, Italy and CNR-Nanotec, Cosenza

*nicolo.maccaferri@uni.lu

*vincenzo.caligiuri@unical.it

## Abstract

Ultrafast control of light-matter interactions constitutes a crucial feature in view of new technological frontiers of information processing. However, conventional optical elements are either static or feature switching speeds that are extremely low with respect to the time scales at which it is possible to control light. Here, we exploit the artificial epsilon-near-zero (ENZ) modes of a metal-insulator-metal nanocavity to tailor the linear photon absorption of our system and realize a non-degenerate all-optical ultrafast modulation of the reflectance at a specific wavelength. Optical pumping of the system at its high energy ENZ mode leads to a strong redshift of the low energy mode because of the transient increase of the local dielectric function, which leads to a sub-3-ps control of the reflectance at a specific wavelength with a relative modulation depth approaching 120%.

## Introduction

Overcoming the fundamental limits of electronics, such as bandwidth, clock-time/frequency and heating of the device, is the main promise of photonics **[1]**. Many recent advancements in this direction rely on the use of light as information carrier, paving the way towards light-based technologies, which will have a huge impact in terms of reduced energy consumption and performance efficiency. Moreover, the possibility of controlling electronics at optical frequencies has recently become possible, thus introducing a new paradigm towards attosecond opto-electronics **[2]**. One way to achieve active control of light is represented by electro-optical modulators **[3]**. Although used in industry **[4]**, they still suffer from limited bandwidth (GHz regime) and large power consumption due to the required electronics. In this framework, it is fundamental to develop new, affordable, and energy-efficient strategies to reach a fast (>100 GHz) and fully tailorable control of optical states at scales which are well below the diffraction limit of electromagnetic radiation. Therefore, all-optical switching [5] has attracted great attention because it can potentially overcome the speed and heat dissipation limitation imposed by electrical switching or passive optical devices [6] [7]. In view of practical applications, the key performing parameters of all-optical switching include modulation depth (defined as the reflection and/or transmission contrast between 'ON' and 'OFF' states) and switching time. The latter defines the bandwidth via the inverse of the transition time between 'ON' and 'OFF' states [5]. Examples of high-speed all-optical switching devices based on semiconductors [8], photonic [9] and plasmonic crystals [10], semiconducting nanostructures [11], metallic [12] and dielectric [13] metasurfaces, single nanoantennas [14] and microring resonators [15] have already been proposed. Another interesting approach is represented by the so-called natural epsilon-near-zero (ENZ) materials [16]. Boltasseva et al. [17] showed that yttrium doped cadmium oxide (CdO) films can enable light intensity switching with relative modulation depths up to 135% in the mid-infrared (mid-IR) region close to the ENZ point and a switching time of 45.6 ps for a pump fluence of 1.3 mJ/cm$^2$. Moreover, Yang et al. [18] have demonstrated that natural ENZ materials can be used for sub-ps switching. Through intra-band optical pumping in an In-doped CdO-based plasmonic perfect absorber, they show an absolute modulation of light intensity of 85.3% with a switching time of 800 fs and using a pump fluence of 0.34 mJ/cm$^2$. In the case of natural ENZ-based switching however, previous approaches are based on the presence of only one ENZ frequency, the exciting polarization can be only transverse magnetic (TM), and very challenging material processing techniques are required to tailor the ENZ wavelength, for instance by material doping. Recently, it has been found that resonances occurring in metal-insulator-metal (MIM) nanocavities can be described as effective ENZ resonances [19]. Several ENZ points, which can be excited with both TM and transverse electric (TE) polarized light, can be designed at will, and their spectral position can be easily engineered by acting on the refractive index and thickness of the embedded dielectric, while their quality(Q)-factor ($\Delta\lambda/\lambda_0$) can be optimized by adopting non-symmetric geometries, yielding low reflectance R at the ENZ modes [20, 21]. Therefore, these systems constitute a promising and flexible alternative to natural ENZ materials.

Here, we propose a novel approach for ultrafast all-optical switching. The basic idea consists in the perturbation of an artificial ENZ symmetric mode in the near-infrared (NIR) through optical pumping of an anti-symmetric one in the ultraviolet (UV) region in a MIM nanocavity. The technological core of the architecture is depicted in **Figure 1**. At the steady state, both the high energy (HE) and the low energy (LE) ENZ modes enable a very high (>90%) photon absorption at the resonances (top panel). The nanocavity is then used for all-optical switching upon photoexcitation. By optical pumping the HE ENZ mode, the LE resonance strongly redshifts because of the transient increase of the dielectric function upon excitation of charge carriers in the metallic layers [22-24], which leads to a modulation of the reflectance R at the wavelength of the LE mode (bottom panel). The non-degenerate approach we propose here allows a strong modulation of the ultrafast nonlinear response of an ENZ mode via linear absorption of the other ENZ mode we pump by using both TM and TE polarized light. It is worth mentioning here that this concept is general, since the cavity resonances can be designed at will in a broad range of wavelengths (from the UV to mid-IR), and it relies on a simple fabrication process (more details are reported in the **Methods** section).

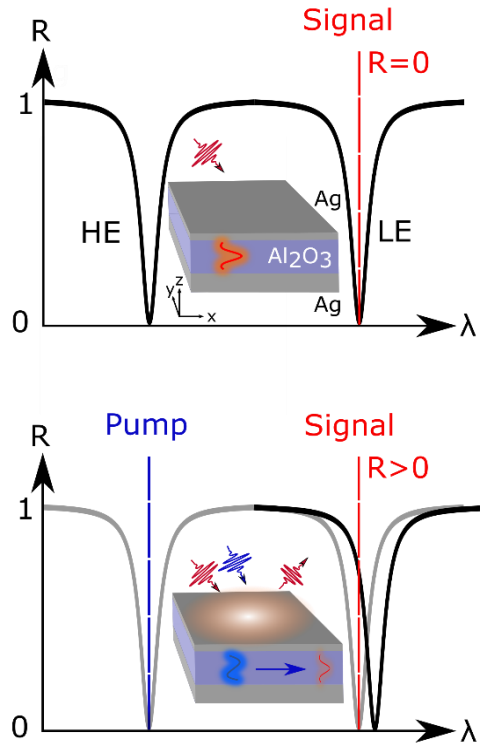

**Figure 1. Ultrafast all-optical switching modulation concept.** Sketch of the all-optical switching concept based on a nanocavity supporting two ENZ modes. Upper panel: Steady-state reflectance R of a MIM nanocavity: both the HE and LE resonances display low R at the ENZ point. Lower panel: Upon optical pumping of the HE ENZ mode, the LE absorption resonance redshifts significantly due to a transient increase of the electronic temperature in the metallic building blocks, which leads to a change of R for the signal probe pulse. The equilibrium positions of the resonances are plotted for reference (grey curve).

## Results

**Steady-state optical response.**

To prove the ENZ nature of our system, we first characterized the steady-state spectral response in terms of absolute R as a function of the angle of incidence θ. In **Figure 2 a,b** we plot the real ($\varepsilon'$, **a**) and imaginary ($\varepsilon''$, **b**) part of the effective dielectric permittivity, as measured by ellipsometry, of the sample consisting of a MIM cavity with Ag[30nm]/Al$_2$O$_3$[180nm]/Ag[100nm] layers (see also **Methods**) for θ = 30° as representative case (for the full angular dependence see **Supplementary Fig. 1**). Noticeably, R shows a pronounced suppression around 327 nm, 395 nm and 730 nm (**Figure 2c**), in correspondence to the three zero-crossings of $\varepsilon'$, thus confirming their ENZ nature. In particular, the mode occurring at 327 nm is the well-known Ferrell-Berreman mode **[25-27]** (labelled FB), while the ones at ~395 nm (labelled HE) and at ~730 nm (labelled LE) correspond to, respectively, the anti-bonding and bonding modes of the MIM nanocavity. In the inset of **Figure 2b** we also plot the near-field profiles of the two ENZ modes calculated via finite element method simulations (details can be found in **Methods**). The LE ENZ mode is the symmetric, and the HE ENZ mode is the anti-symmetric mode of the MIM cavity **[20]**. Interestingly, both ENZ modes can be excited with similar efficiency using either transverse electric (TE, s-polarized) or transverse magnetic (TM, p-polarized) for the incoming light (see also **Supplementary Fig. 1**), showing that the ENZ resonances represent photonic, rather than plasmonic modes of the cavity. This functionality represents an additional feature if we compare our artificial ENZ nanocavity to natural ENZ media that only support ENZ resonances for p-polarized incident light.

To confirm the validity of the measured effective permittivities, simulations based on the Transfer Matrix Method (TMM) have been performed by considering light impinging at 30° on one homogenized layer with thickness equal to that of the MIM and dielectric permittivity ε' and ε''. The corresponding R (**Figure 2c,** red dashed curve) is in excellent agreement with both experiments (**Figure 2c,** white circles) and classic layer-by-layer (**Figure 2c,** black solid curve) simulations.

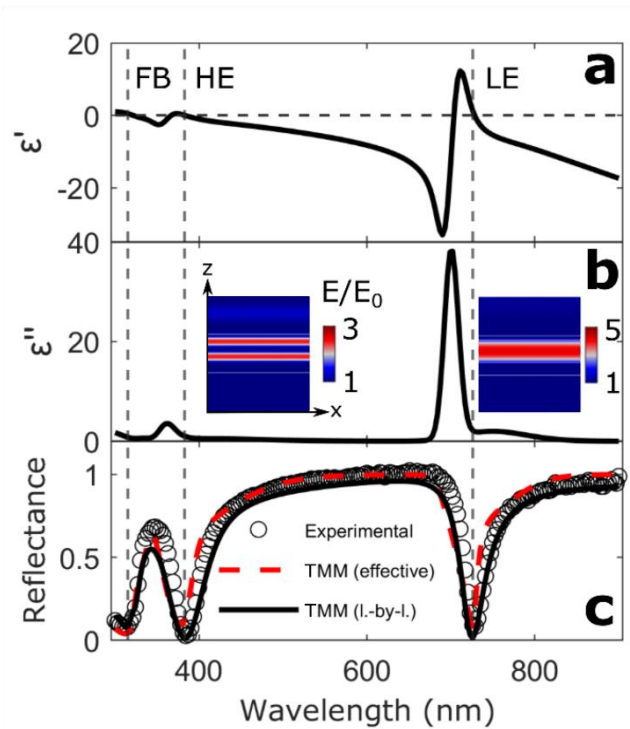

**Figure 2. Steady state response of the MIM nanocavity.** Steady-state characterization of the MIM nanocavity made of Ag[30nm]/Al2O3[180nm]/Ag[100nm] layers on a glass substrate as measured at 30° angle of incidence. Real (a) and imaginary (b) parts of the dielectric permittivity of the MIM nanocavity measured by spectroscopic ellipsometry. The insets in (b) show the profiles of the electric field amplitude (color code) of the HE and LE ENZ modes normalized to the incoming electric field amplitude $E_0$ in dependence of spatial coordinates x and z. (c) Experimental (white circles) reflectance compared to TMM calculations carried out by considering a single layer with the measured effective dielectric permittivity (red dashed line) shown in (a) and (b) and the real (layer-by-layer) structure (black solid curve).

**All-optical modulation of the nonlinear optical response of the nanocavity.**

All-optical modulation of the reflectance of the system has been proved by performing wavelength- and time-resolved pump-probe experiments. The MIM nanocavity was pumped at the HE ENZ mode, while the temporal dynamics were investigated by probing in the 715 nm -760 nm spectral range (**Figure 3a**), where the LE ENZ mode is located (more details on the pump and probe signals, as well as on the experimental setup **[28]**, can be found in **Methods**, **Supplementary Fig. 1** and **Supplementary Fig. 3**). An incident angle of 30° was chosen for the ultrafast experiments, but our approach can in principle be generalized also to other angles due to the preserved high Q-factor over the angular dispersion of the ENZ modes (see **Supplementary Fig. 1).** Upon resonant pumping of the HE mode, electrons are photoexcited in the metallic layers, and quickly thermalize via electron-electron scattering, leading to an elevated electronic temperature and thus a transient increase of the local dielectric function [23]. This introduces a red-shift of the LE ENZ resonance due to the local increase of permittivity, and thus a pump-induced change ΔR/R close to the LE resonance. As can be inferred by **Figure 2c**, we decided to pump the HE mode since it is the one showing the largest photon absorption (97 %). Moreover, the HE mode has also a larger width, so that any tiny shift of this mode is less appreciable compared to the shift of the LE mode which features a higher quality factor. Therefore, we use the LE mode as the mode we want to modulate the nonlinear response, because it is the one that physically can display the largest variation. We do not exclude that, by pumping directly the LE mode, we might obtain even larger shifts, as recently demonstrated in a multi-layered structure. However, this is against the idea to use the non-degenerate pump-probe scheme we are proposing in this work.

For a pump fluence of ~5.2 mJ/cm$^2$, we indeed observe a positive change of the reflectance ΔR around 10 % at the wavelength of the LE ENZ mode, corresponding to a modulation ΔR/R of about 120%, which is our best result so far, and a negative ΔR/R of about 50% at longer wavelengths (above 735 nm), as it can be seen in **Figure 3b** and **Figure 4a**. The induced reflectance modulations, either positive or negative, are strongly localized at specific wavelengths, and do not shift throughout the relaxation process, at least within the first 5 ps. The absorption of incident light by the metallic layers is drastically enhanced by the presence of the HE mode, thus driving efficient carrier excitation.

The physics underlying the dynamics of the nanocavity is dominated by the photoexcitation of charge carriers in the constituent metallic building blocks. In the visible (VIS) and NIR spectral range, the steady-state optical properties of silver can be well described using the Drude model [29]

$$\varepsilon = \varepsilon_{\infty,0} - \frac{\omega_p^2}{\omega(\omega + i\gamma_0)} \qquad (1)$$

In this model, $\gamma_0$ is the damping rate at the equilibrium, and the plasma frequency is given by $\omega_p^2 = \frac{ne^2}{m_{\text{eff}}\varepsilon_0}$, where $n$ is the electron density, $e$ the electron charge, $m_{\text{eff}}$ the effective electron mass and $\varepsilon_0$ the vacuum permittivity. Contributions to $\varepsilon$ from interband transition of d band electrons to the sp band can empirically be accounted for by adding functions of Lorentz oscillators, where the sum is represented in the steady state by $\varepsilon_{\infty,0}$. For $\gamma_0 \ll \omega$, Equation (2) can be decomposed into contributions to the real $\varepsilon'$ and imaginary part $\varepsilon''$ of the permittivity as

$$\varepsilon' = \varepsilon_{\infty,0} - \frac{\omega_{p,0}^2}{\omega^2}, \qquad \varepsilon'' = \frac{\gamma_0 \omega_{p,0}^2}{\omega^3}. \qquad (2)$$

The impulsive intraband excitation of electrons in the conduction band initially creates an non-thermal distribution of electrons, which quickly equilibrates mainly via electron-electron and electron-phonon scattering, leading to an elevated electronic temperature accompanied by drastic changes in the interband contribution to $\varepsilon$ [22].

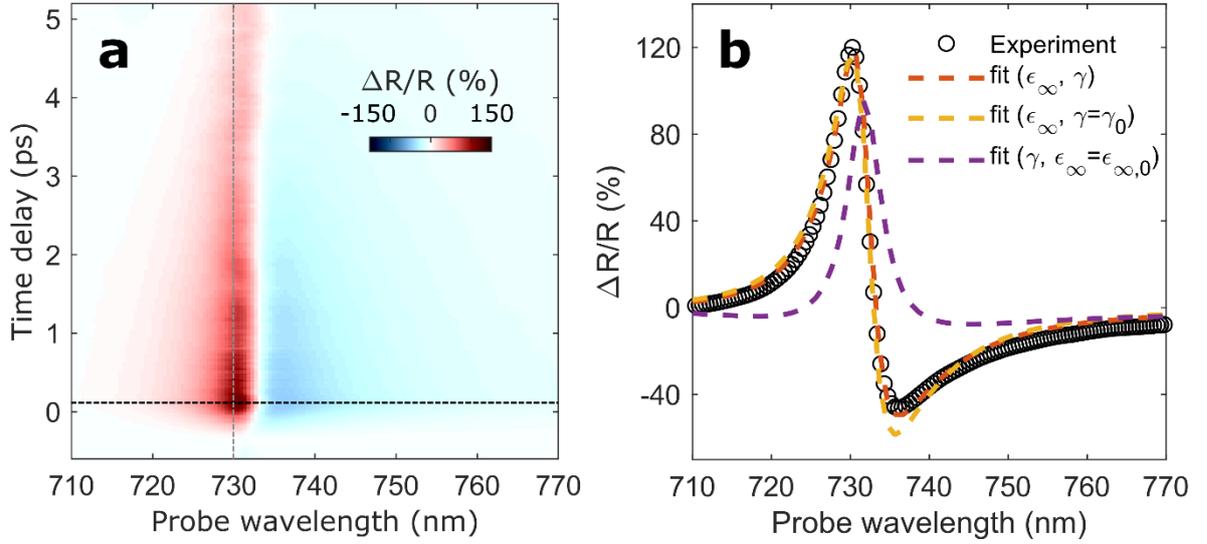

**Figure 3. Ultrafast modulation of the photon reflectance.** Time-resolved modulation of R at the LE resonance. (a) Time-resolved ΔR/R pump-probe spectrum in the range 710 nm – 770 nm for a pump fluence of 5.2 mJ/cm². The equilibrium LE ENZ mode (vertical dashed line, grey) strongly redshifts upon pumping of the HE ENZ mode, thus enabling an overall modulation of R exceeding 120%. (b) Cut along the spectrum (white circles) shown in (a) at 100 fs time delay (horizontal dashed line in (a), black). Dashed lines show fits to the data based on multilayer TMM simulations, where the damping $\gamma$ (violet), the interband term $\varepsilon_\infty$ (yellow), or both $\varepsilon_\infty$ and $\gamma$ (orange) of the silver permittivity are changed.

More in detail, the hot electron temperature induced variation of the electronic energy distribution in the conduction band leads to a reduction/increase of the state occupation probability below/above the Fermi energy. That results in a complex modification of the interband transition probability, that can be computed if the band structure of the material is known [30], and leads to a complex behaviour of $\varepsilon_\infty$ around the interband transition threshold (300 nm in silver) [31]. However, $\varepsilon_\infty$ is only weakly dispersive in the spectral range of the LE resonance, i.e. at 730 nm, due to the much higher energy of the interband absorption edge (in this case $\varepsilon_\infty$ reduces to a constant contribution to the permittivity). Thus, the transient changes of $\varepsilon_\infty$ can assumed to be almost independent of the wavelength in the spectral region that is probed in this experiment, leading to a constant offset $\Delta\varepsilon_\infty = \varepsilon_\infty - \varepsilon_{\infty,0}$ of the steady-state permittivity. To fit the experimental ΔR/R spectrum at 100 fs time delay (see **Figure 3 b**), we use the Drude-Lorentz parameters given by Rakic et.al [32] for silver in the steady state and vary the value of $\Delta\varepsilon_\infty$ in TMM-based simulations. We find a very good agreement with the experimental ΔR/R data for $\Delta\varepsilon_\infty = 0.5$, indicating that the transient increase of $\varepsilon_\infty$ is central for the observed reflectance modulation. Also, changes in $\gamma$ due to an elevated electronic temperature are considered. By selectively changing $\varepsilon_\infty$ and $\gamma$ in the fit, the dominant role of $\varepsilon_\infty$ is further evidenced. When both, $\gamma$ and $\varepsilon_\infty$ are varied, we obtain a slightly better fit and observe a transient increase of $\gamma$, $\Delta\gamma = \gamma - \gamma_0 = 6$ meV, that can be attributed to an increased electron–electron and electron-phonon coupling for higher electronic temperature. Let us note that this analysis emphasizes the dominant role of an increased $\varepsilon'$ for the observed redshift of the LE mode. As can be inferred by Equation 2, $\varepsilon'$ also changes with the plasma frequency $\omega_p$. Indeed, if small deviations from a perfectly parabolic shape of the conduction band in silver are taken into account, also $\omega_p$ changes with the electronic temperature, due to a transient modification of the electron effective mass $m_{\text{eff}}$. Using our simple approach, contributions coming from changes in either $\omega_p$ or $\varepsilon_\infty$ cannot be fully decoupled. Although it is widely established that changes in $\omega_p$ are more related to the lattice temperature [33][34], our quantitative modelling depends strongly on that assumption.

Finally, to evaluate the temporal dynamics and magnitude of the switching process, we fitted the time evolution of ∆R/R at the position of the LE resonance, i.e. at 730 nm. A single-exponential relaxation model at time delays larger than 200 fs yields the precise decay time of the modulation. Hereby, we also include a constant offset, reflecting dynamics much longer than the temporal window investigated in the experiments. In a second step, we fit a sophisticated model to the complete dynamics between -0.5 ps and 5 ps based on a Gaussian error function in combination with a double-exponential decay and constant offset, where we fix the slower time constant to the decay time obtained in the first step. From this, we extract the maximum modulation while ensuring the stability of the fit (see **Figure 4a** and **Methods**).

**Performance and limits.**

The overall performance of our system is characterized by a series of measurements where we varied the excitation pulse fluence from 1.0 mJ/cm$^2$ up to 7.0 mJ/cm$^2$, shown in **Figure 4a**. We find an upper working threshold of the device at ~5.2 mJ/cm$^2$ (see **Figure 4b**).

At high fluence (>5.2 mJ/cm$^2$), we observe a large decrease of ∆R/R along with irreversible sample damage (see black circles in **Figure 4b** and also **Supplementary Fig. 5**). However, this is not a limiting aspect, since in practical applications lower pulse energies are more relevant than relative modulation depths exceeding 100%, in particular in terms of energy consumption. The relaxation of the system is a complex interplay of several processes triggered by the optical pump that induces a non-equilibrium distribution of the electrons within the metal. The subsequent thermalization of hot electrons with the lattice is the dominant relaxation mechanism [22]. Additional scattering with acoustic phonons and heat diffusion within the metallic films [24] also influence the time scales observed in the experiments. The overall measured relaxation time will thus be determined by the time scales, relative contributions and the interplay of these mechanisms, which all depend on excitation fluence. However, as can be seen in **Figure 4b**, the decay time remains approximately 3 ps in the range of investigated fluences. For a pump fluence of 5.2 mJ/cm$^2$, we find a decay time of $(2.5 \pm 0.3)$ ps corresponding to an all-optical switching bandwidth of about 400 GHz with a relative modulation depth of about 120%. In our case, the switching time is intrinsically limited to the ps time scale by the carrier density and electron heat capacity of the metals that are used. The bottleneck of the device is thus the switching time rather than the spectral bandwidth (>10 nm) of operation, which is dictated by the Q-factor of the ENZ modes. In principle, to overcome this limitation, by combining layers of transparent conducting oxides (TCOs) and dielectrics, the device would benefit from the lower carrier density and electron heat capacity of TCOs as compared to the noble metals [35]. This would translate to a faster electron-phonon coupling and thus enable even higher switching speeds of the cavity. Moreover, the MIM nanocavity can be engineered to work at a desired wavelength, from UV to mid-IR, with practically no limitation regarding the materials involved and the free spectral range between the ENZ resonances can be easily engineered by exploiting multiple cavity geometries, without affecting the Q-factor of the resonances **[21]**.

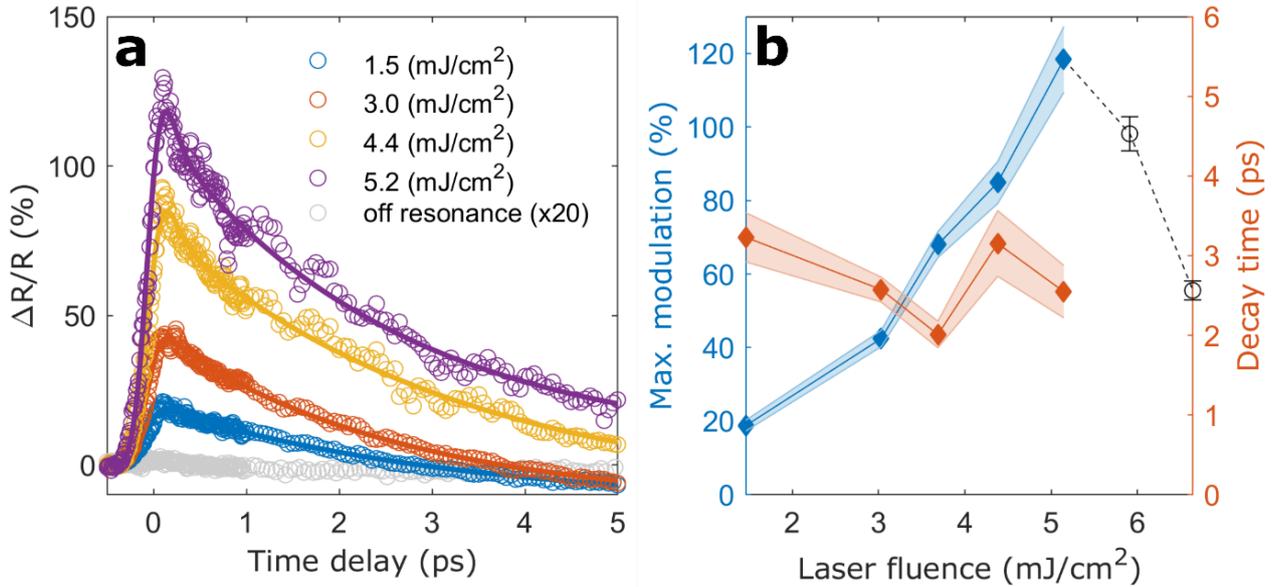

**Figure 4. Performances and limits of the MIM nanocavity-based all-optical switching.** Switching performance of our device for different pump pulse fluences. (a) Time-dependent ΔR/R at probe wavelength $\lambda$ = 730 nm for increasing excitation pulse fluences from 1.0 mJ/cm$^2$ up to 7.0 mJ/cm$^2$. The grey circles correspond to off-resonant pumping at $\lambda$ = 610 nm with pump fluence of 3.3 mJ/cm$^2$. Solid lines show a double-exponential fit model as described in the main text. (b) Maximum modulation and decay time as a function of pump pulse fluence. Maximum modulation was extracted from a double-exponential fit, while the decay time was obtained by a single-exponential fit considering only time delays larger than 200 fs. Shaded areas indicate the fit uncertainty. Below ~5.0 mJ/cm$^2$ (blue diamonds) we find ultrafast switching behaviour, yielding outstanding modulation of about 120 % at 5.2 mJ/cm$^2$. At higher fluence (black circles), the signal modulation decreases rapidly due to irreversible sample damage. The decay time does not change substantially and it is around 3 ps for all measurements.

Finally, it is worth mentioning here another interesting effect observed in our experiments. After the initial electronic relaxation (time delays > 5 ps), oscillations of ΔR/R on the order of 10 % can be observed within the time-delay range we can explore in our experiments (see **Supplementary Fig. 4**). After arrival of the pump pulse, part of the optical energy is converted into mechanical energy due to photoinduced thermal stress. This leads to the formation of acoustic shockwaves and thus a transient opto-acoustic modification of the sample reflectance. In more detail, after Fourier analysis of the signal at the wavelength of the LE mode, fast (~60 GHz) and slow (~10 GHz) oscillations can be distinguished, corresponding to the propagation of acoustic shockwaves in the metal and dielectric layers respectively. The induced modulations then damp away in a few hundred ps via attenuation of the optically triggered acoustic shockwaves (see also **Supplementary Fig. 4b**). Although these results go beyond the scope of the current work, we do not exclude the possibility that by engineering the acoustic response of our cavity, namely by creating a hybrid opto-acoustic cavity, we could match the oscillation frequency of an acoustic mode of the cavity in the GHz range, thus potentially enhance the opto-acoustic modulation of the nonlinear optical response of our system beyond the observed 10%. This additional functionality might be then further explored to achieve a multi-functional device which combine optical and acousto-mechanical properties, which can be also eventually modulated by using external agents such as magnetic fields or spin currents if the metallic layers are made of a magnetic material [**36**,**37**].

## Discussion

We have demonstrated all-optical, ultrafast (sub-3-ps) switching of the reflectance of a metal-insulator-metal nanocavity approaching a relative modulation depth of 120% in the VIS-NIR spectral range. Our approach is based on the high absorbance of the nanocavity ENZ modes, whose spectral position can easily be tailored at will from UV to mid-IR frequencies, thus lifting from demanding fabrication processes to tailor the spectral position of the ENZ resonance. Via pumping of one ENZ mode, we achieve a relative modulation of reflectance at wavelengths close to the other mode. Without the need of driving higher order effects for ultrafast switching, our system is based on linear absorption, providing large relative modulation exceeding 100% and switching bandwidths of few hundred GHz at moderate excitation fluence, due to the high Q-factor of the ENZ modes. Moreover, the proposed system can work with both TE and TM polarization, which is not the case for natural ENZ materials, so our system is also more versatile if then it will be implemented in practical photonics applications after being optimized and better engineered, which are goals out of the scope of the present manuscript. Our approach proves that we can design at will a tailorable pumping channel and thus a tunable probe signal modulation, which is not easily achievable in natural ENZ materials, where only doping can shift the ENZ point. Furthemore, the primary benefit in terms of implementation in real-world devices is that these structures are easier to fabricate or integrate with other systems and that they have a response that is more widely tunable lying on the simple argument that the ENZ resonance positions depend on the geometry, while their physical limitation on the materials that are used, which are also well established and readily available. Finally, we foresee that this approach can also be used as platform for the ultrafast manipulation of optical nonlinearities, such as second and third harmonic generation, Purcell factor enhancement and other very promising future and emerging light-driven technologies.

# Methods

**Fabrication.**

Metal-insulator-metal samples have been prepared by electron beam evaporation in a custom-made vacuum chamber at the base pressure of 1x10$^{-6}$ mbar. Ag and Al$_2$O$_3$ layers have been deposited at 0.2 and 0.4 Å/sec, respectively. The layer thicknesses have been measured by quartz microbalance.

**Optical characterization.**

Steady state optical response of the samples has been recorded with a V-VASE J.A. Woollam spectroscopic ellipsometer. Spectroscopic ellipsometry supplied with p- and s-polarized transmittance and reflectance in the spectral range between 300 and 1300 nm was performed to measure the ellipsometric angles ψ and Δ, whose fitting led to the effective dielectric permittivity of the sample. P- and s-polarization R and transmittance measurements were performed in the angular range from 30° to 80°.

**Pump-probe experiments.**

Transient reflection measurements are carried out with a home built spectroscopy system based on a commercial Yb:KGW regenerative amplifier system at a laser repetition rate of 50 kHz **[28]**. A non-collinear optical parametric amplifier (NOPA) working in the VIS/NIR spectral range initially delivers bandwidth-limited pulses at 790 nm that are frequency-doubled using a BBO crystal yielding the final pump pulses with 1.8 nm spectral bandwidth (FWHM) centred at 395 nm (Fourier limit of pulses ~130 fs). Residual spectral components at lower energy are suppressed with a dielectric short-pass filter (Thorlabs, FESH500). For the off-resonant pumping experiments, another NOPA working in the VIS is used, where a 610 nm band-pass filter (Edmund Optics) is put before the parametric amplification, yielding final pump pulses with 8 nm spectral bandwidth centred at 610 nm (Fourier limit of pulses ~70 fs). The pump induced change of reflection is probed by a white-light supercontinuum between 500 nm and 900 nm which is temporally compressed by custom designed dielectric chirped mirrors (DCMs). The probe pulse energy is then adjusted to ensure a 1:20 energy ratio compared to the pump. Using an off-axis parabolic (OAP) mirror with 50.8 mm focal length, focal diameters of 20 µm and 25 µm are achieved for probe and pump pulses respectively. Pump and probe pulses are focused onto the sample non-collinearly, in order to spatially filter the probe pulse after sample interaction. Residual scattered pump radiation is further spectrally suppressed with a dielectric long-pass filter (Thorlabs, FELH600). Spectrally resolved detection of the probe pulse after sample interaction is achieved by using a spectrograph (Acton) in combination with a high-speed charge coupled device (CCD) camera operating at 50 kHz. Finally, a Pockels cell modulates the pump pulse train at half the repetition rate of the laser system, allowing the calculation of ΔR/R on a 25 kHz basis.

**Two-step fit model.**

To extract the decay time of the modulation, the following model 1. was used.

1. Model decay time $\tau$: $M(t) = A \cdot \exp\left(-\frac{t-t_0}{\tau}\right) + C$ (for $t > 0.2$ ps)

The amplitude of the modulation is found as the maximum of the following model 2., that describes the complete dynamics from -0.5 ps to 5 ps. The time constant $\tau$ is fixed in this model and taken from model 1.

2. Model amplitude with fixed $\tau$: $M(t) = \frac{1}{2}\left[\text{erf}\left(\frac{t-t_0}{\sigma}\right) + 1\right] \cdot \left[B_{\text{fast}} \exp\left(-\frac{t-t_0}{\tau_{\text{fast}}}\right) + B \exp\left(-\frac{t-t_0}{\tau}\right) + C_0\right]$

**Simulations.**

Numerical simulations were performed with the finite elements method using the commercial COMSOL Multiphysics software. The geometry was set up in 2D and periodic boundary conditions (PBCs) with Floquet-periodicity were used for the simulation. Linearly polarized light (plane wave), either with p- or s-polarization, at different wavelengths was generated via periodic ports at the top of the simulation geometry. Interpolated data from Rakic et.al **[32]** and from Boidin et al. **[38]** were used to describe the linear optical properties of the silver and alumina layers, respectively. Transfer Matrix Method based simulations have been carried out via a custom MatLAB code based on the classic formulation that can be found in ref **[39].**

## Acknowledgements

N.M. acknowledges support from the Luxembourg National Research Fund (CORE Grant No. C19/MS/13624497). D.B. acknowledge support of the Deutsche Forschungsgemeinschaft through the Emmy Noether programme (BR 5030/1-1) and support by the European Research Council through grant number 819871 (UpTEMPO). D.B. and N.M. acknowledge support from the FEDER Program (grant n. 2017-03-022-19 Lux-Ultra-Fast). V.C., A.D.L. and R.K. acknowledge support from the Project TEHRIS, that is part of ATTRACT, that has received funding from the European Union's Horizon 2020 Research and Innovation Programme under Grant Agreement No. 777222.

## Author contributions

J.K., V.C. and N.M. conceived and developed the concept. J.K. and J.A. performed the time-resolved pump-probe experiments. V.C. designed the structure with inputs from J.K. and N.M., and D.G. fabricated the samples and characterized the steady state optical response. V.C. performed numerical simulations and semi-analytical description of the steady-state epsilon-near-zero behaviour of the nanocavity. J.K. developed the model describing the ultrafast dynamics. J.K., D.B. and N.M. analysed the data. N.M. supervised the work. All the authors participated in the discussion and in the manuscript preparation.

# Supplementary Material

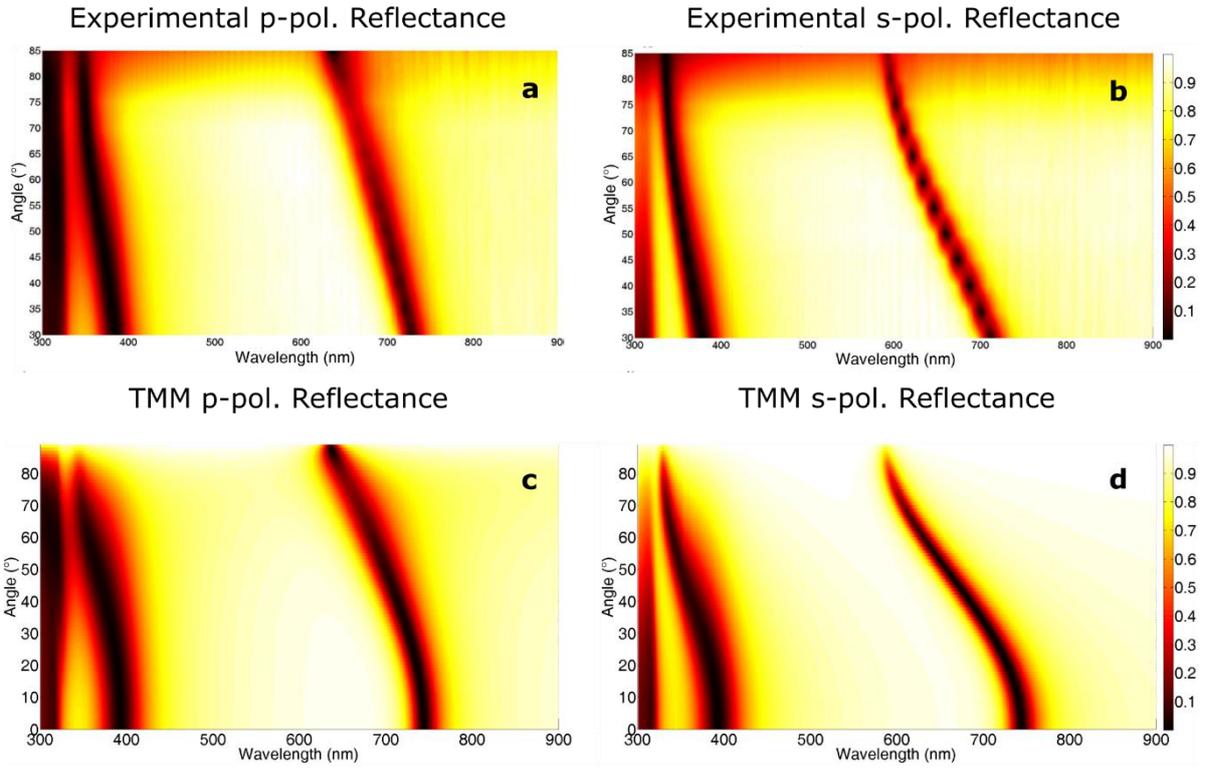

**Supplementary Figure 1.** Angular and wavelength dependence of the steady state R of the MIM nanocavity for s- and p-polarization of the incident light. (a) and (b) show experimental data and (c) and (d) show layer-by-layer TMM calculations for an extended angular range.

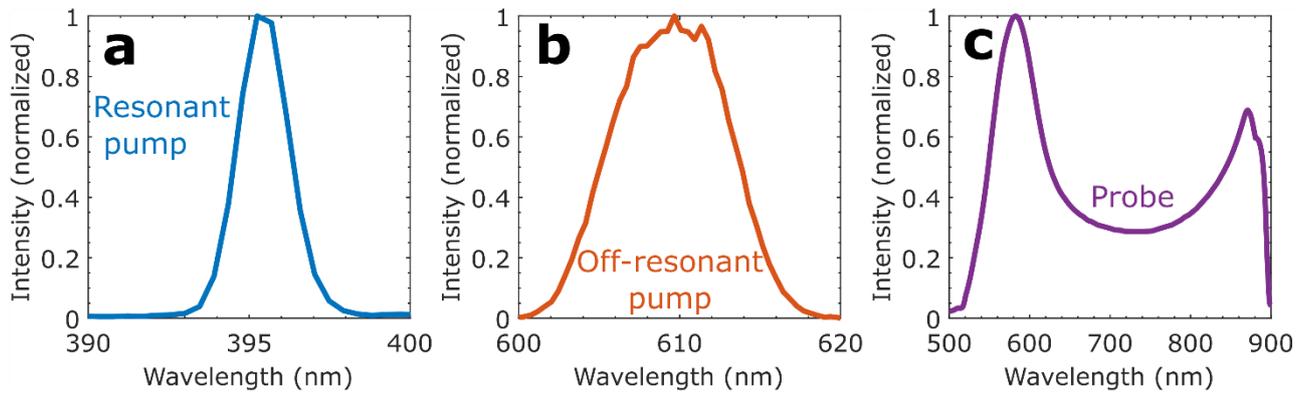

**Supplementary Figure 2.** Normalized spectra of the resonant pump (a), off-resonant pump and probe (b) pulses.

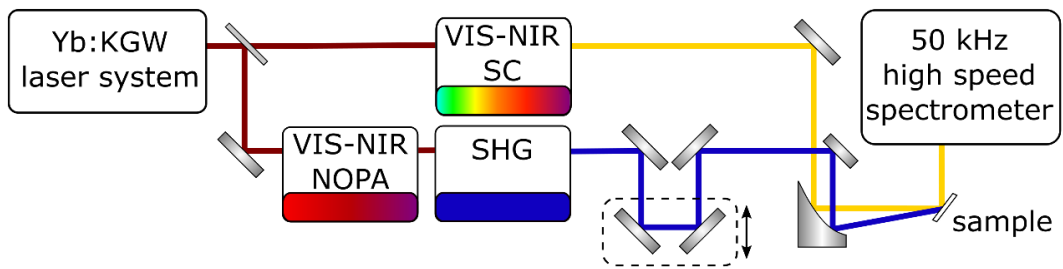

**Supplementary Figure 3.** Sketch of the experimental setup used in the ultrafast all-optical switching experiments.

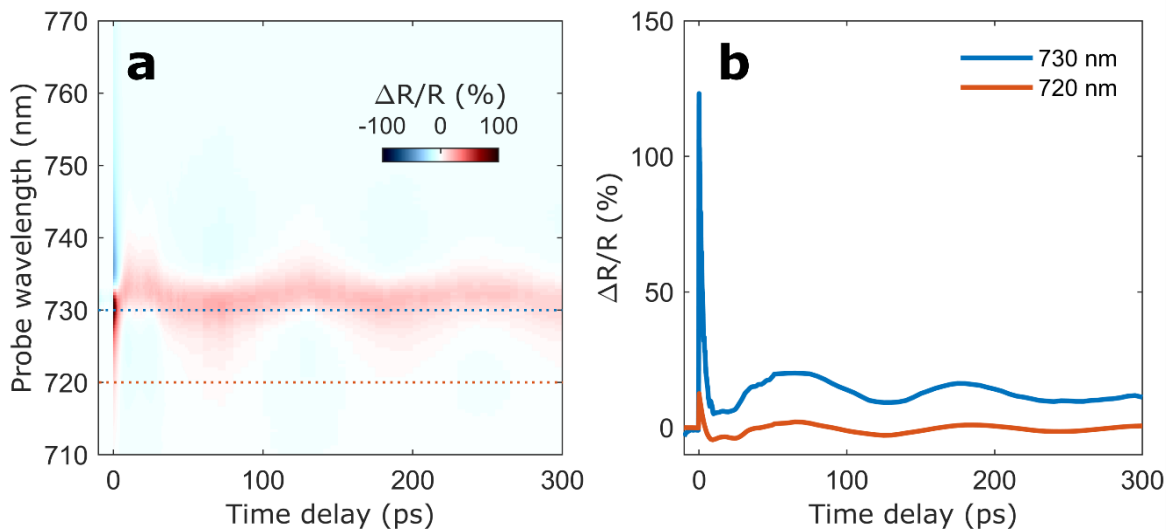

**Supplementary Figure 4.** Opto-acoustic modulation of R. (a) Time-resolved ΔR/R pump-probe spectrum for the measurement shown in **Figure 3a** in the main text for an extended range of the time delay up to 300 ps. After the ultrafast electronic relaxation within the first 5 ps, ΔR/R shows the development of acoustic shockwaves due to the intense pump pulse. (b) Cuts of the spectrum shown in (a) at the wavelength of the LE resonance (730 nm, blue curve) and at lower wavelength (720 nm, red curve).

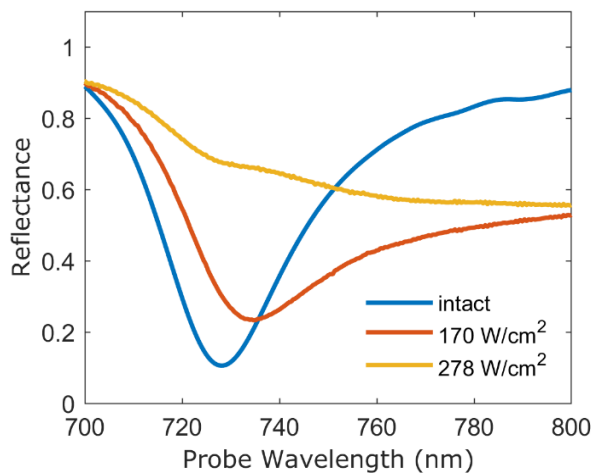

**Supplementary Figure 5.** Linear reflectance of the sample after high-intensity pumping for several minutes. Red and yellow curves correspond to the linear optical response after a pump-probe measurement with a pump fluence of 6.8 mJ/cm$^2$ and 11.0 mJ/cm$^2$, thus confirming permanent damage to the sample for high pump fluence.

# References

[1] International Roadmap for Devices and Systems https://irds.ieee.org/roadmap-2017 (2017).

[2] M. Ludwig, G. Aguirregabiria, F. Ritzkowsky, T. Rybka, D. Codruta Marinica, J. Aizpurua, A. G. Borisov, A. Leitenstorfer, and D. Brida, "Sub-femtosecond electron transport in a nanoscale gap," Nat. Phys. 16, 341–345 (2020).

[3] G. T. Reed, G. Mashanovich, F. Y. Gardes and D. J. Thomson "Silicon optical modulators," Nat. Photon. 4, 518-526 (2010).

[4] M. Smit et al. "An introduction to InP-based generic integration technology," Semicond. Sci. Technol. 29 083001 (2014).

[5] Z. Chai, X. Hu, F. Wang, X. Niu, J. Xie, and Q. Gong, "Ultrafast All-Optical Switching," Adv. Optical Mater. 5, 1600665 (2017).

[6] K. Bergman, J. Shalf, T. Hausken, "Optical interconnects and extreme computing," Opt. Photonics News 27, 32–39 (2016)

[7] S. Werner, J. Navaridas, and M. Luján, "A survey on optical network-on-chip architectures," ACM Comput. Surv. 50, 1–37 (2017).

[8] G. Grinblat, M. P. Nielsen, P. Dichtl, Y. Li, R. F. Oulton, and S. A. Maier, "Ultrafast sub–30-fs all-optical switching based on gallium phosphide," Sci. Adv. 5, eaaw3262 (2019).

[9] K. Nozaki, T. Tanabe, A. Shinya, S. Matsuo, T. Sato, H. Taniyama, and M. Notomi, "Sub-femtojoule all-optical switching using a photonic-crystal nanocavity," Nat. Photon. 4, 477–483 (2010).

[10] M. Pohl, V. I. Belotelov, I. A. Akimov, S. Kasture, A. S. Vengurlekar, A. V. Gopal, A. K. Zvezdin, D. R. Yakovlev, and M. Bayer, "Plasmonic crystals for ultrafast nanophotonics: Optical switching of surface plasmon polaritons," Phys. Rev. B 85, 081401(R) (2012).

[11] M. P. Fischer, C. Schmidt, E. Sakat, J. Stock, A. Samarelli, J. Frigerio, M. Ortolani, D. J. Paul, G. Isella, A. Leitenstorfer, P. Biagioni, and D. Brida, "Optical activation of germanium plasmonic antennas in the mid-infrared," Phys. Rev. Lett. 117, 047401 (2016).

[12] M. Ren, B. Jia, J.-Y. Ou, E. Plum, J. Zhang, K. F. MacDonald, A. E. Nikolaenko, J. Xu, M. Gu, and N. I. Zheludev, "Nanostructured plasmonic medium for terahertz bandwidth all-optical switching," Adv. Mat. 23, 5540–5544 (2011).

[13] M. R. Shcherbakov, S. Liu, V. V. Zubyuk, A. Vaskin, P. P. Vabishchevich, G. Keeler, T. Pertsch, T. V. Dolgova, I. Staude, I. Brener, and A. A. Fedyanin, "Ultrafast all-optical tuning of direct-gap semiconductor metasurfaces," Nat. Commun. 8, 17 (2017).

[14] G. Grinblat, R. Berté, M. P. Nielsen, Y. Li, R. F. Oulton, and S. A. Maier, "Sub-20 fs All-Optical Switching in a Single Au-Clad Si Nanodisk," Nano Lett. 18, 7896-7900 (2018).

[15] J. S. Pelc, K. Rivoire, S. Vo, C. Santori, D. A. Fattal, and R. G. Beausoleil, "Picosecond all-optical switching in hydrogenated amorphous silicon microring resonators," Opt. Express, 22, 3797-3810 (2014).

[16] N. Kinsey, C. Devault, J. Kim, M. Ferrera, V. M. Shalaev, and A. Boltasseva, "Epsilon-near-zero Al-doped ZnO for ultrafast switching at telecom wavelengths," Optica 2, 616–622 (2015).

[17] S. Saha, B. T. Diroll, J. Shank, Z. Kudyshev, A. Dutta, S. N. Chowdhury, T. S. Luk, S. Campione, R. D. Schaller, V. M. Shalaev, A. Boltasseva, and M. G. Wood, "Broadband, High-Speed, and Large-Amplitude Dynamic Optical Switching with Yttrium-Doped Cadmium Oxide," Adv. Funct. Mater. 1908377 (2019).

[18] Y. Yang, K. Kelley, E. Sachet, S. Campione, T. S. Luk, J.-P. Maria, M. B. Sinclair, and I. Brener, "Femtosecond optical polarization switching using a cadmium oxide-based perfect absorber," Nat. Photon. 11, 390–395 (2017).

[19] V. Caligiuri, M. Palei, M. Imran, L. Manna, and R. Krahne, "Planar Double-Epsilon-Near-Zero Cavities for Spontaneous Emission and Purcell Effect Enhancement," ACS Photon. 5, 2287–2294 (2018).

[20] V. Caligiuri, M. Palei, G. Biffi, S. Artyukhin, and R. Krahne, "A Semi-Classical View on Epsilon-Near-Zero Resonant Tunneling Modes in Metal/Insulator/Metal Nanocavities," Nano Lett. 19, 3151–3160 (2019).

[21] V. Caligiuri, M. Palei, G. Biffi, and R. Krahne, "Hybridization of epsilon-near-zero modes via resonant tunneling in layered metal-insulator double nanocavities," Nanophotonics 8, 1505-1512 (2019).

[22] G. Della Valle, M. Conforti, S. Longhi, G. Cerullo, and D. Brida, "Real-time optical mapping of the dynamics of nonthermal electrons in thin gold films," Phys. Rev. B 86, 155139 (2012).

[23] N. Del Fatti, C. Voisin, M. Achermann, S. Tzortzakis, D. Christofilos, and F. Vallée, "Nonequilibrium electron dynamics in noble metals," Phys. Rev. B 61, 16956 (2000).

[24] J. Hohlfeld, S. S. Wellershoff, J. Güdde, U. Conrad, V. Jähnke, and E. Matthias, "Electron and lattice dynamics following optical excitation of metals," Chem. Phys. 251, 237-258 (2000).

[25] R. A. Ferrell, and E. A. Stern, "Plasma resonance in the electrodynamics of metal films," Am. J. Phys. 30, 810–812 (1962).

[26] R. A. Ferrell, "Predicted Radiation of Plasma Oscillations in Metal Films," Phys. Rev. 111, 1214-1222 (1958).

[27] D. W. Berreman, "Infrared Absorption at Longitudinal Optic Frequency in Cubic Crystal Films," Phys. Rev. 130, 2193-2198 (1963).

[28] A. Grupp, A. Budweg, M. P. Fischer, J. Allerbeck, G. Soavi, A. Leitenstorfer, and D. Brida, "Broadly tunable ultrafast pump-probe system operating at multi-kHz repetition rate," J. Opt. 20, 014005 (2017).

[29] M. I. Marković, and A. D. Rakić, "Determination of optical properties of aluminium including electron reradiation in the Lorentz-Drude model," Opt. Laser Technol. 22, 394-398 (1990).

[30] R. Rosei "Temperature modulation of the optical transitions involving the Fermi surface in Ag: Theory" Phys. Rev. B 10, 474-483 (1970).

[31] J.-Y. Bigot, V. Halté, J.-C. Merle, and A. Daunois "Electron dynamics in metallic nanoparticles," Chem. Phys. 251, 181-203 (2000).

[32] A. D. Rakić, A. B. Djurišic, J. M. Elazar, and M. L. Majewski, "Optical properties of metallic films for vertical-cavity optoelectronic devices," Appl. Opt. 37, 5271-5283 (1998).

[33] D. T. Owens, C. Fuentes-Hernandez, J. M. Hales, J. W. Perry, and B. Kippelen "A comprehensive analysis of the contributions to the nonlinear optical properties of thin Ag films," J. Appl. Phys. 107, 123114 (2010).

[34] H. Baida, D. Mongin, D. Christofilos, G. Bachelier, A. Crut, P. Maioli, N. Del Fatti, and F. Vallée "Ultrafast nonlinear optical response of a single gold nanorod near its surface plasmon resonance," Phys. Re. Lett. 107, 057402 (2011).

[35] D. Traviss, R. Bruck, B. Mills, M. Abb, and O. L. Muskens "Ultrafast plasmonics using transparent conductive oxide hybrids in the epsilon-near-zero regime," Appl. Phys. Lett. 102, 121112 (2013).

[36] V. V. Temnov "Ultrafast acousto-magneto-plasmonics" Nat. Photon. volume 6, 728–736 (2012).

[37] I. Razdolski, A. Alekhin, N. Ilin, J. P. Meyburg, V. Roddatis, D. Diesing, U. Bovensiepen, and A. Melnikov "Nanoscale interface confinement of ultrafast spin transfer torque driving non-uniform spin dynamics," Nat. Commun. 8, 15007 (2017).

[38] R. Boidin, T. Halenkovič, V. Nazabal, L. Beneš, and P. Němec, "Pulsed laser deposited alumina thin films," Ceram. Int. 42, 1177-1182 (2016).

[39] M. Born, and E. Wolf "Principles of Optics Electromagnetic Theory of Propagation, Interference and Diffraction of Light." Pergamon Press (1980).